\documentclass[a4paper,11pt]{article}

\usepackage{jcappub} 

\usepackage[T1]{fontenc} 

\usepackage{units}
\usepackage[figuresright]{rotating}
\usepackage{graphicx}
\usepackage{xspace}
\usepackage{units}
\usepackage{xcolor}
\usepackage{subfig}
\usepackage{multirow}
\usepackage{cleveref}

\def\eV{\ifmmode {\mathrm{\ e\kern -0.1em V}}\else
\textrm{e\kern -0.1em V}\fi\xspace}%

\newcommand{\hess}{{\scshape H.E.S.S.}\xspace}

\title{An upper limit on the cosmic-ray luminosity of individual sources from gamma-ray observations}

\author[a]{A.D. Supanitsky,}
\author[b]{V. de Souza}

\affiliation[a]{Instituto de Astronom\'{\i}a y F\'{\i}sica del Espacio (IAFE), CONICET-UBA, Argentina}
\affiliation[b]{Instituto de F\'{\i}sica de S\~{a}o Carlos, Universidade
de S\~ao Paulo, Brazil}

\emailAdd{supanitsky@iafe.uba.ar}
\emailAdd{vitor@ifsc.usp.br}

\abstract{Different types of extragalactic objects are known to
  produce TeV gamma-rays. Some of these objects are the most probable
  candidates to accelerate cosmic rays up to $10^{20}$ eV.  It is very
  well known that gamma-rays can be produced as a result of the
  cosmic ray propagation through the intergalactic medium.  These
  gamma-rays contribute to the total flux observed in the direction of
  the source. In this paper we propose a new method to derive an upper
  limit on the cosmic-ray luminosity of an individual source based on
  the measured upper limit on the integral flux of GeV-TeV
  gamma-rays. We show how it is possible to calculate an upper limit
  on the cosmic-ray luminosity of a particular source and we explore
  the parameter space in which the current GeV-TeV gamma-ray
  measurements can offer a useful determination. We study in detail
  two particular sources, Pictor A and NGC 7469, and we calculate the
  upper limit on the proton luminosity of each source based on the
  upper limit on the integral gamma-ray flux measured by the \hess
  telescopes.}

\begin{document}
\maketitle
\flushbottom

\section{Introduction}
\label{sec:intro}

Individual sources of ultra high energy cosmic rays
(UHECR)\footnote{Hereafter cosmic rays will refer to the hadronic
component of the total flux. Photons and neutrinos can be part of
the flux observed at Earth but they are a minority component.} with
energy above $10^{18}$ eV have not been identified yet due to the severe
limitations imposed by the uncertainties on the determination of the
identity of the particles and the unknown strength and structure of
the magnetic fields in the Universe. The reconstructed direction of a
charged cosmic-ray particle might not point back exactly to its source
due to deviation in the magnetic fields. Given this limitation, the
study of the origin of the cosmic rays is restricted to limited
classes of sources and any important information involves the
convolution of measured quantities with propagation models and
catalogs of objects~\cite{bib:Auger:CenA}.

One possible exception to this scenario which needs extra data to be
confirmed is Cen A. The Pierre Auger Observatory has measured an
excess of events around Cen A \cite{bib:Auger:CenA}, however the
correlation angle is very large ($\sim 18$ degrees) and many other
objects lie within this angular window.

On the other hand, gamma-rays are not deviated by magnetic
fields. Cosmic-ray observatories are sensitive to gamma-rays at the
highest energies ($E_\gamma \gtrsim 10^{18}$ eV). At present, no
observation of a high energy photon has been claimed, all candidate
events are compatible with nuclei primaries. As a consequence upper
limits on the gamma-ray flux at the highest energies have been
obtained~\cite{bib:UpperLim}. At TeV energies a few hundred gamma-ray
sources have been discovered by the past and current Imaging
Atmospheric Cherenkov Telescopes (IACTs). Also, at lower energies (GeV)
more than one thousand gamma-ray sources have been detected by the Fermi-LAT
instrument~\cite{bib:Fermi}. It is believed that the majority of the
gamma-rays observed in the GeV-TeV energy range originate in those
sources. However a significant number of these gamma-rays can be produced
in the propagation of cosmic rays in the intergalactic
medium~\cite{bib:GammaCR:1,bib:GammaCR:2,bib:GammaCR:3,bib:GammaCR:4,bib:GammaCR:5}.

UHECRs interact with the radiation backgrounds producing secondary\footnote{The
name primary is used for particles (cosmic rays and gamma-rays) produced in the
sources. Here, the word secondary names particles produced on the way from the
source to Earth.} particles. The production of secondary particles in the cosmic 
microwave background and in the extragalactic background light has been studied 
in detail (see for instance \cite{bib:pp:1,bib:pp:2,bib:pp:3}). The relevant aspect 
of the propagation of UHECR for this paper is the production of GeV-TeV gamma-rays 
which can be summarized as follows.

Pions, electron-positron pairs, and neutrons are the main outcome of
the interaction of the primary cosmic ray with the radiation fields.
For the considered distances (extragalactic sources) pions and neutrons
decay before reaching Earth. Gammas, electrons and neutrinos are the
most numerous secondary particles produced in the cascade process.
Gamma-rays ($E \gtrsim 10$ GeV) interact with the low energy photons
of the radiation fields producing electron-positron pairs. Electrons
and positrons can upscatter low energy photons of the background via
inverse Compton producing gamma-rays that continue the cascade process.

The relevant photon backgrounds for this cascading process are: (a)
The extragalactic background light composed of the infrared, visible,
and ultraviolet regions, which is important in the energy range from
$E_{\gamma} = 10$ GeV to $E_{\gamma} = 10^5$ GeV of the incident
gamma, (b) the cosmic microwave background from $E_{\gamma} = 10^5$
GeV to $E_{\gamma} = 10^{10}$ GeV, and (c) the radio background from
$E_{\gamma} = 10^{10}$ GeV to $E_{\gamma} = 10^{13}$ GeV
\cite{bib:DeAngelies:13}.

The cascade process generates a high flux of secondary photons with
energy in the GeV-TeV range. Due to the production of secondary
gamma-rays in the propagation of cosmic rays in the intergalactic
medium, the number of gamma-rays measured at any given direction in
the sky is a combination of the ones produced in the source (primary
flux) and the ones produced during the propagation of the cosmic rays
(secondary flux). This is an ambiguity which has to be solved in order
to extract the primary gamma-ray spectrum of a source. In fact, it has
been pointed out that the measured gamma-ray spectra of some unusually
bright distant blazars might originate as a result of the propagation
of high energy protons emitted by these sources
\cite{bib:GammaPr:1,bib:GammaPr:2,bib:GammaPr:3}.

Among all candidate sources observed by the ground and space based observatories,
there are some for which only an upper limit on the gamma-ray flux is
obtained. For these objects the upper limit constrains the sum of the primary and
secondary fluxes of photons. If the measured upper limit flux of GeV-TeV
gamma-rays is interpreted as the upper limit on the secondary gamma-rays produced
by the cosmic rays propagating from a distance source to Earth, it is possible to set
an upper limit on the cosmic-ray luminosity of the source. If the primary gamma-ray
flux of the source is not taken into account then the calculated upper limit
on the cosmic-ray luminosity is the most conservative.

The method proposed here relies on calculations of the UHECR propagation in
the intergalactic medium to establish the connection between the UHECR and GeV-TeV
gamma-rays fluxes. The method can be applied to any upper limit on the GeV-TeV
gamma-ray flux measured by ground or space based instruments. In order to
illustrate the method we have used a catalog of sources for which the \hess
telescopes have measured upper limits~\cite{bib:hess}.

Two particular sources Pictor A and NGC 7469 are studied in
detail. Pictor A is a Fanarof-Riley II galaxy with high radio
activity. Its jet is clearly visible by
Chandra~\cite{bib:pictor:a:chandra} and it is oriented at a large
angle with respect to Earth, therefore the gamma-ray emission is not
expected to be high. The \hess telescopes observed Pictor A for 7.9 h,
no signal was detected and an upper limit on the integral flux of gamma-rays was
obtained. NGC 7469 is a Seyfert 1 spiral galaxy placed at $\sim 68$
Mpc from the Earth. It has been observed by \hess for 3.4 h, no excess
was seen and an upper limit on the integral flux of gamma-ray was also calculated.
The method developed in this paper allows us to calculate an upper
limit on the proton luminosities of Pictor A and NGC 7469, for energies
above $10^{18}$ eV, based on the upper limits on the integral flux of
gamma-ray obtained by \hess.

The new method proposed here has its relevance increased by the proposed
construction of the CTA Observatory~\cite{bib:cta}. The CTA Observatory
will have a sensitivity one order of magnitude better than the current
IACTs and therefore is going to be able to set very strict upper limits
at GeV-TeV energies.

At the same time, the scientific community is planning the next steps
for the detection of UHECR~\cite{bib:jem:euso,bib:auger:next}. Multi-messenger
astroparticle physics has been claimed to be the key to solving the remaining
puzzles in high energy astrophysics~\cite{bib:multi:1,bib:multi:2,bib:multi:3}.
The idea presented in this paper contributes in this direction.

\section{The method}
\label{sec:method}

The method developed in this section establishes the connection between a measured
upper limit on the integral flux of GeV-TeV gamma-rays of a source and the UHECR
cosmic-ray luminosity of the same source. The purpose of this section is to formalize 
the method in a general mathematical way. The details of the application of the method 
to specific sources are done in section~\ref{sec:data}.

Let us start by assuming that a point source injects a cosmic ray energy spectrum
given by a power law with an exponential cutoff
\begin{equation}
\frac{dN}{dE dt}=\frac{L_{CR}}{C_0}\ E^{-\alpha} \exp(-E/E_{cut}),
\label{eq:InjSpec}
\end{equation}
where $L_{CR}$ is the cosmic-ray luminosity, $\alpha$ is the spectral
index, $E_{cut}$ is the cutoff energy, and $C_0$ is a normalization
constant
\begin{equation}
C_0 =  \int_{E_{min}}^{\infty} dE \ E^{-\alpha+1} \exp(-E/E_{cut}),
\label{eq:C0}
\end{equation}
where $E_{min} = 10^{18}$ eV is the minimum energy of a cosmic-ray particle
produced by the source.

The injection spectrum can be rewritten as
\begin{equation}
\frac{dN}{dE dt}=\frac{L_{CR}}{\langle E \rangle_0}\ P_{CR}^0(E),
\label{eq:InjSpecNorm}
\end{equation}
where $P_{CR}^0(E)$ is the normalized energy distribution of the injected particles
\begin{equation}
P_{cr}^0(E) = \frac{ E^{-\alpha} \exp(-E/E_{cut}) } { \mathop{\displaystyle \int_{E_{min}}^{\infty} dE \ E^{-\alpha} \exp(-E/E_{cut})} },
\end{equation}
and $\langle E \rangle_0$ is the mean energy
\begin{equation}
\langle E \rangle_0 = \int_{E_{min}}^{\infty} dE \ E\ P_{CR}^0(E) =
\frac{ \mathop{\displaystyle \int_{E_{min}}^{\infty} dE \ E^{-\alpha+1} \exp(-E/E_{cut}) }     }%
{ \mathop{\displaystyle \int_{E_{min}}^{\infty} dE \ E^{-\alpha} \exp(-E/E_{cut}) }   }.
\label{eq:MeanE}
\end{equation}

On the way from the source to Earth, cosmic ray particles interact
with radiation fields causing a modification of the energy
distribution from $P_{CR}^0(E)$ to $P_{CR}(E)$. For a source at a given
comoving distance ($D_s$) from Earth, assuming an isotropic emission,
the cosmic-ray flux can be written as
\begin{equation}
I_{CR}(E) = \frac{L_{CR}}{4 \pi D_s^2\ (1+z_s) \langle E \rangle_0 }\ K_{CR}\ P_{CR}(E),
\label{eq:CRFlux}
\end{equation}
where  $z_s$ is the redshift of the source, $P_{CR}(E)$ is the energy
distribution of particles arriving on Earth and $K_{CR}$ is the number
of cosmic rays arriving on Earth per injected particle. $K_{CR}$ and
$P_{CR}(E)$ include the physics of the propagation of the cosmic rays,
i.e.~all interactions and energy losses suffered during the propagation
process. Here an isotropic cosmic ray emission is assumed.

This source generates a secondary gamma-ray flux which is proportional
to its cosmic-ray luminosity. In analogy to equation~(\ref{eq:CRFlux})
it is possible to write the gamma-ray flux at Earth as a function of
the cosmic-rays luminosity,
\begin{equation}
I_{\gamma}(E_\gamma) = \frac{L_{CR}}{4 \pi D_s^2\ (1+z_s) \langle E \rangle_0 }\ K_{\gamma}\ P_{\gamma}(E_\gamma),
\label{eq:GammaFlux}
\end{equation}
where $K_\gamma$ is the number of gamma-rays generated per injected
cosmic-ray particle and $P_\gamma(E_\gamma)$ is the energy distribution
of the gamma-rays arriving on Earth. Here $E_\gamma$ is used to
denote the energy of the gamma-rays. Note that in this case $K_\gamma$
and $P_\gamma(E_\gamma)$ include the physics of the propagation of the
cosmic-rays and gamma-rays.

Equation~(\ref{eq:GammaFlux}) relates the gamma-ray flux measured on
Earth to the cosmic-ray luminosity. Therefore, an upper limit on the
integral gamma-ray flux can be translated into an upper limit on the
cosmic-ray luminosity ($L_{CR}^{UL}$) of the source according to
\begin{equation}
L_{CR}^{UL} = I_{\gamma}^{UL}(>E_\gamma^{th})\ \frac{4 \pi
D_s^2\ (1+z_s) \langle E \rangle_0 }{\mathop{\displaystyle%
\ K_{\gamma} \  \int_{E_{\gamma}^{th}}^\infty%
dE\ P_{\gamma}(E_\gamma)}},
\label{eq:CRUL}
\end{equation}
where $I_{\gamma}^{UL}(>E_{\gamma}^{th})$ is the measured upper limit
on the integral gamma-ray flux obtained above a given energy threshold
($E_{\gamma}^{th}$) at a given confidence level (CL).

Equation~(\ref{eq:CRUL}) is the summary of the method presented here. It
allows us to calculate an upper limit on the cosmic-ray luminosity given
the measured upper limit on the integral gamma-ray flux. $K_{\gamma}$ and
$P_{\gamma}(E_\gamma)$ summarize the propagation process under the assumption
of the source properties: distance ($D_S$) and injection spectrum
($P_{CR}^0(E)$).

In general, the factors $K_{CR}$ and $K_{\gamma}$ and the distribution functions
$P_{CR}(E)$ and $P_{\gamma}(E_\gamma)$ can be obtained from Monte Carlo
simulations~\cite{bib:crpropa} (see Appendix~\ref{app:sim} for details).

It is worth noting that when a mixture of nuclei are injected by the source
such that the proton fraction is different from zero, equation (\ref{eq:CRUL})
when calculated for proton primaries (using $P_\gamma^{pr}$ and $K_\gamma^{pr}$)
gives an upper limit on the proton luminosity of the source. This can be
understood from the fact that, if heavier nuclei are injected by the source in
addition to proton primaries, the resulting gamma-ray flux at Earth is larger than
the one expected for protons only.

\section{Application of the method}
\label{sec:data}

The data considered in this work were taken by the \hess
telescopes~\cite{bib:hess}. \hess is an important ground-based
gamma-ray observatory currently in operation that makes use of the
imaging atmospheric Cherenkov technique. Between 2005 and 2007, \hess
observed several active galactic nuclei from which a gamma-ray flux
with energy above 100 GeV was expected to be measured.  At this
energy, \hess's sensitivity is at the level of a few percent of the
Crab Nebula flux.

In Ref.~\cite{bib:hess:meas} the \hess Collaboration reported the
observation of 14 candidate sources for which no significant emission
was detected. From the 14 candidate sources, 10 were Blazar and 4
Non-Blazar types of AGNs. In that publication the upper limit on the
integral gamma-ray flux of each source was reported. We select three
sources from the \hess publication to illustrate the method proposed
here: NGC 1068, NGC 7469, and Pictor A. These three sources were chosen
by a combination of factors. They are all non-Blazar, they cover the
distance range of interest ($\sim16$, $\sim68$, and $\sim141$ Mpc), and
they are all in the field of view of the Pierre Auger Observatory.

The factor $K_{\gamma}$ and the energy distribution $P_{\gamma}(E_\gamma)$
for each of the three sources were calculated by using the Monte Carlo simulation
program CRPropa~\cite{bib:crpropa} (see Appendix~\ref{app:sim} for details)
under the following input assumptions. The injection spectrum is a power law
with an exponential cutoff as shown by equation~(\ref{eq:InjSpec}). $E_{cut}$
is taken to be proportional to the charge number ($Z$) of the injected particle
$E_{cut}^Z = Z \times E_{cut}^{pr}$, where $pr = $ proton. Two cases were simulated:
Only proton and only iron nuclei as primary particles. The following cutoff energies
were considered: $\log( E_{cut}^{pr}/ \textrm{eV})  = 20, \; 20.25, \;
20.5, \; 20.75$ and  $21$ eV and the corresponding $E_{cut}^{26}$ for iron nuclei.
For each combination of the injected particle type and $E_{cut}$ nine values of the
spectral index ($\alpha$), going from 2 to $2.8$ in steps of 0.1, were considered.

Using the upper limit on the integral flux of gamma-rays
($I_{\gamma}^{UL}(>E_{\gamma}^{th})$) obtained by \hess the upper limit on the
cosmic-ray luminosity $L_{CR}^{UL}$ is calculated, by mean of equation~(\ref{eq:CRUL}),
for each simulated scenario. The upper limit UHECR flux obtained for each source by
using equation (\ref{eq:CRFlux}) is therefore determined by the injection spectrum
shape ($\alpha$ and $E_{cut}$) and by the calculated $L_{CR}^{UL}$.

The upper limit on the cosmic-ray luminosity of a given point source calculated by
using this method is useful only if the contribution of the corresponding UHECR flux
to the total flux observed by a given cosmic-ray observatory, which includes all sources
in its field of view, is smaller than the upper limit on the UHECR total flux measured
by this observatory and calculated at the same CL. Because a given point source contributes
in a different way to the total UHECR flux observed by different cosmic-ray observatories,
this condition has to be fulfilled considering the observations done by every cosmic-ray
observatory relevant in this energy range.

The contribution to the spectrum measured by the Pierre Auger Observatory of the upper
limit UHECR flux calculated for NGC 1068, for all scenarios considered, does not fulfill
the condition of being lower than the upper limit on the UHECR total flux measured by the
Pierre Auger Observatory. This indicates that the upper limit measured by \hess is not
restrictive enough in order to be used by this method. In this sense, the spectrum measured
by cosmic-ray experiments can be used as a classification of the restriction power of the
GeV-TeV gamma-ray measurements. For the cases of NGC 7469 and Pictor A this condition is
fulfilled for all proton scenarios considered. These sources are analyzed in detail in the
next subsections.

Note that Pictor A is only seen by the Pierre Auger Observatory. However NGC 7469 and NGC 1068
are also seen by Telescope Array which is located in the northern hemisphere. In particular,
NGC 7469 has a positive declination and is better seen by Telescope Array than by Auger.
However, for these two sources it can be seen that the restriction power of the data used to
estimate the surface detector spectrum of Telescope Array~\cite{Ivanov:12} is smaller than the
one corresponding to the Pierre Auger Observatory, i.e.~the upper limit on the total flux obtained
from the surface detector data of Telescope Array is larger than the one corresponding to Auger.
This is due to essentially two reasons, the flux observed by Telescope Array is larger than the
one observed by Auger and the statistics of Auger is larger than the one corresponding to Telescope
Array. Therefore, for the analyses of NGC 7469 and Pictor A only the Auger spectrum is considered.

\subsection{Pictor A}
\label{sec:pictor:a}

Pictor A is an interesting Fanarof-Riley II broad emission line radio
galaxy~\cite{bib:pictor:a:1,bib:pictor:a:2}. Due to its relative
proximity ($z=0.0342$) it has been extensively observed in many
wavelength ranges. Radio observations have shown a relativistic jet
with rich hot spots in the termination shock region~\cite{bib:pictor:a:chandra}.
Recently, Fermi-LAT has detected GeV gamma-rays (0.2 < E < 300 GeV) with a very low
flux ($F_{0.2 < E < 300 \textrm{GeV}} = (5.8 \pm 0.7) \times 10^{-9} \; \; cm^{-2} \; \;
s^{-1}$) coming from this source~\cite{bib:pictor:a:3}.

The abundance of data from this source (see Ref.~\cite{bib:pictor:a:review:1} for
a compilation) and the possibility of resolving its inner structures make it an
important object with which to probe to many of the unsolved questions regarding
gamma-ray emission. The broad band emission of Pictor A suggests a synchrotron
self-Compton mechanism operating in the source in which TeV gamma-rays can be
produced via inverse-Compton scattering. However the predicted flux is well below
the measurements of Fermi-LAT and \hess~\cite{bib:pictor:a:3}.

Regarding the acceleration of the highest energetic particles, Pictor A
is also a plausible candidate. A magnetic field of the order of $1\ \mu$Gauss
has been measured in filaments with kpc size~\cite{bib:pictor:a:4} which
guarantees the possibility of accelerating particles up to $E \sim 10^{20}$
eV according to the Hillas criteria~\cite{bib:hillas:criteria}. At the same
time, the measured luminosity of the jet $L_{jet} \sim 10^{43.1}$ erg s$^{-1}$
also satisfies the Lovelace~\cite{bib:lovelace} condition for acceleration in
the jet.

The upper limit on the integral gamma-ray flux of Pictor A obtained by \hess, at 99.9\% CL, is
$I_\gamma^{UL}(E>E_\gamma^{th}=320\ \textrm{GeV}) = 2.45\times 10^{-12}$ cm$^{-2}$ s$^{-1}$. As
explained above, by using this upper limit and equation~(\ref{eq:CRUL}), the upper limit on
the cosmic-ray luminosity of Pictor A is calculated for all scenarios considered. Then, the upper
limit on the UHECR flux is calculated by using the upper limit on the cosmic-ray luminosity.

\begin{figure}[h]
  \centering
  \subfloat[Pr $- E_{cut} = 10^{20}$ eV]{\includegraphics[width=0.5\textwidth]{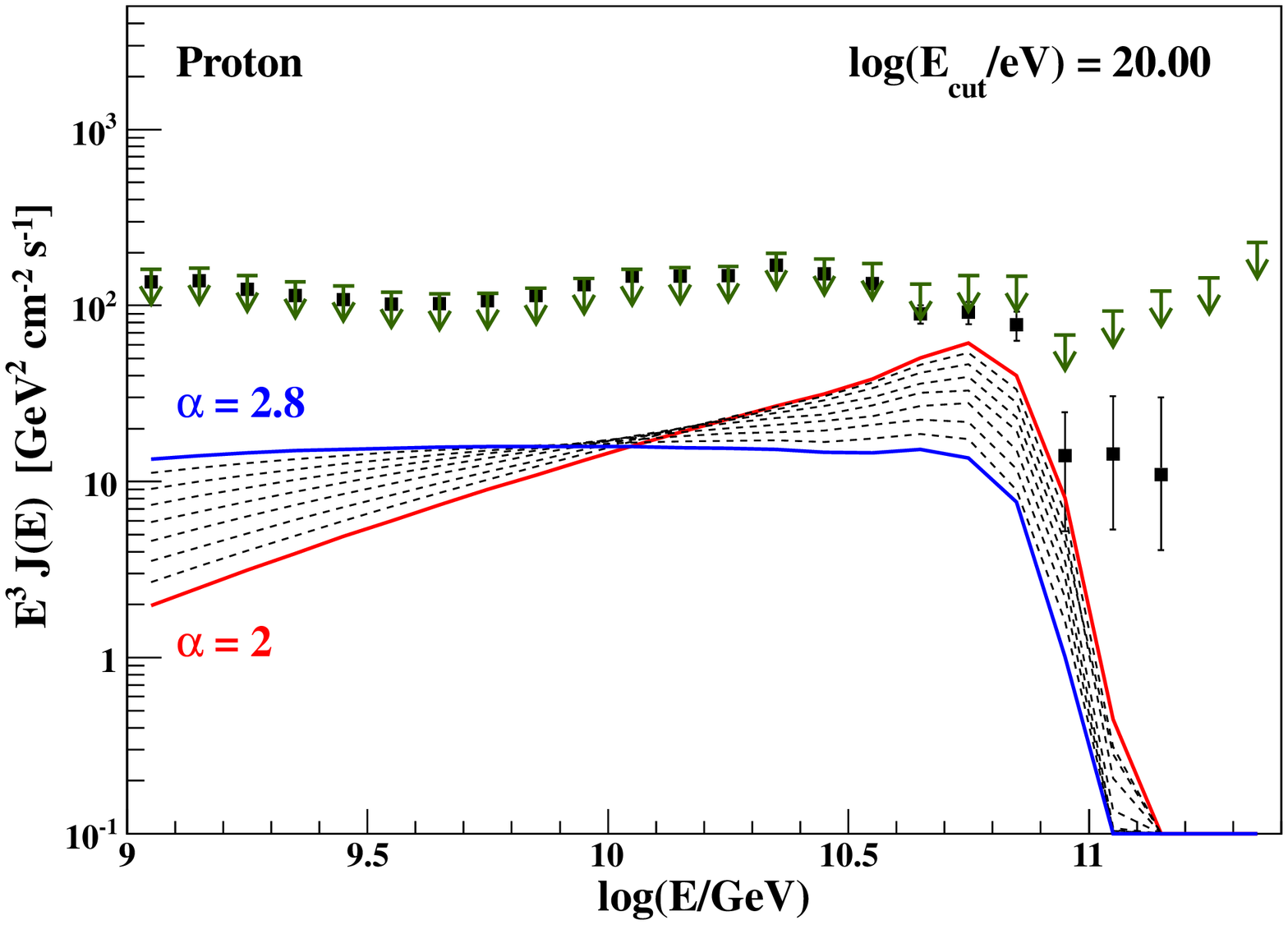}}
  \subfloat[Fe $- E_{cut} = 26\times10^{20}$ eV]{\includegraphics[width=0.5\textwidth]{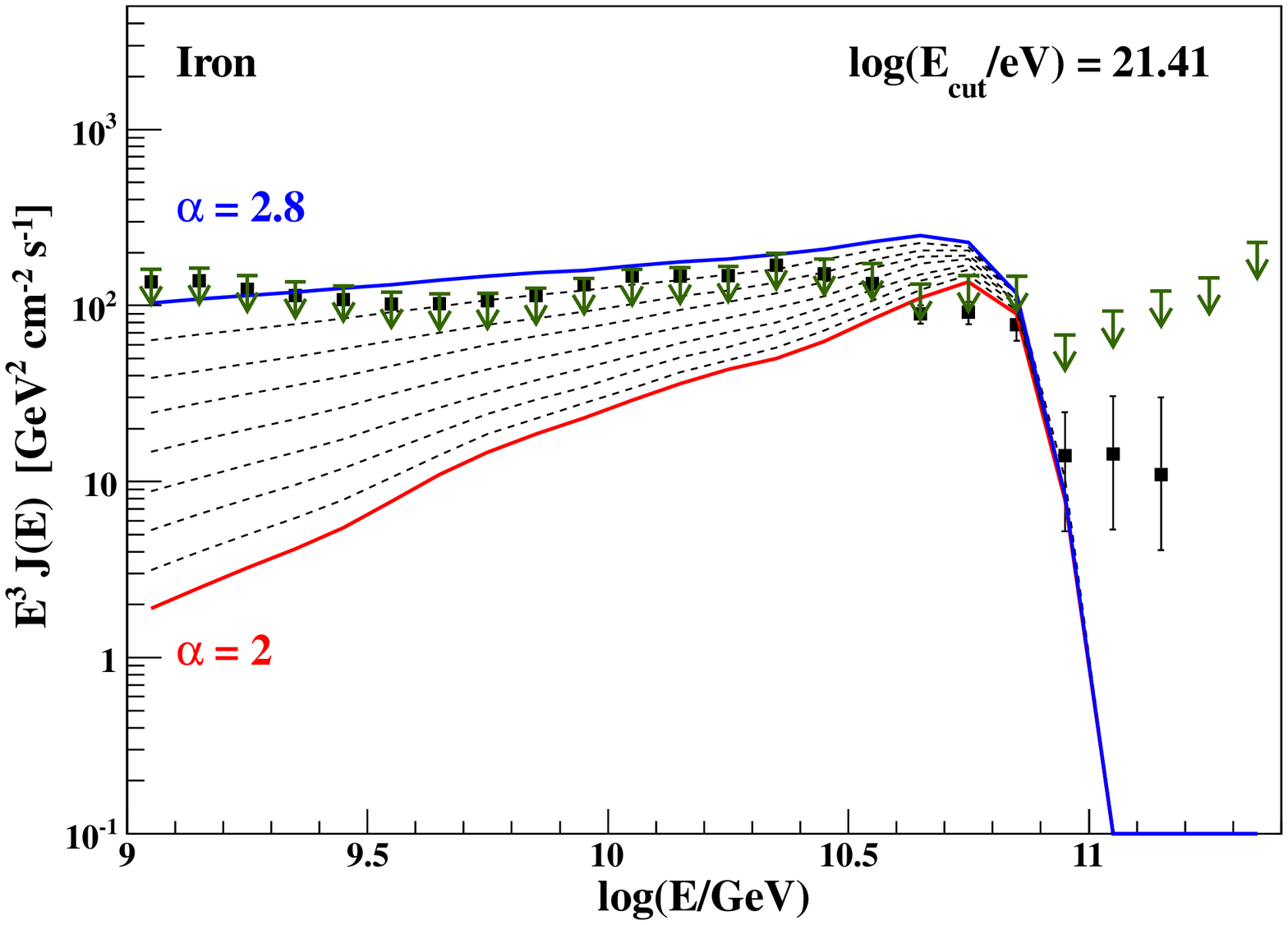}}\\

  \subfloat[Pr $- E_{cut} = 10^{20.5}$ eV]{\includegraphics[width=0.5\textwidth]{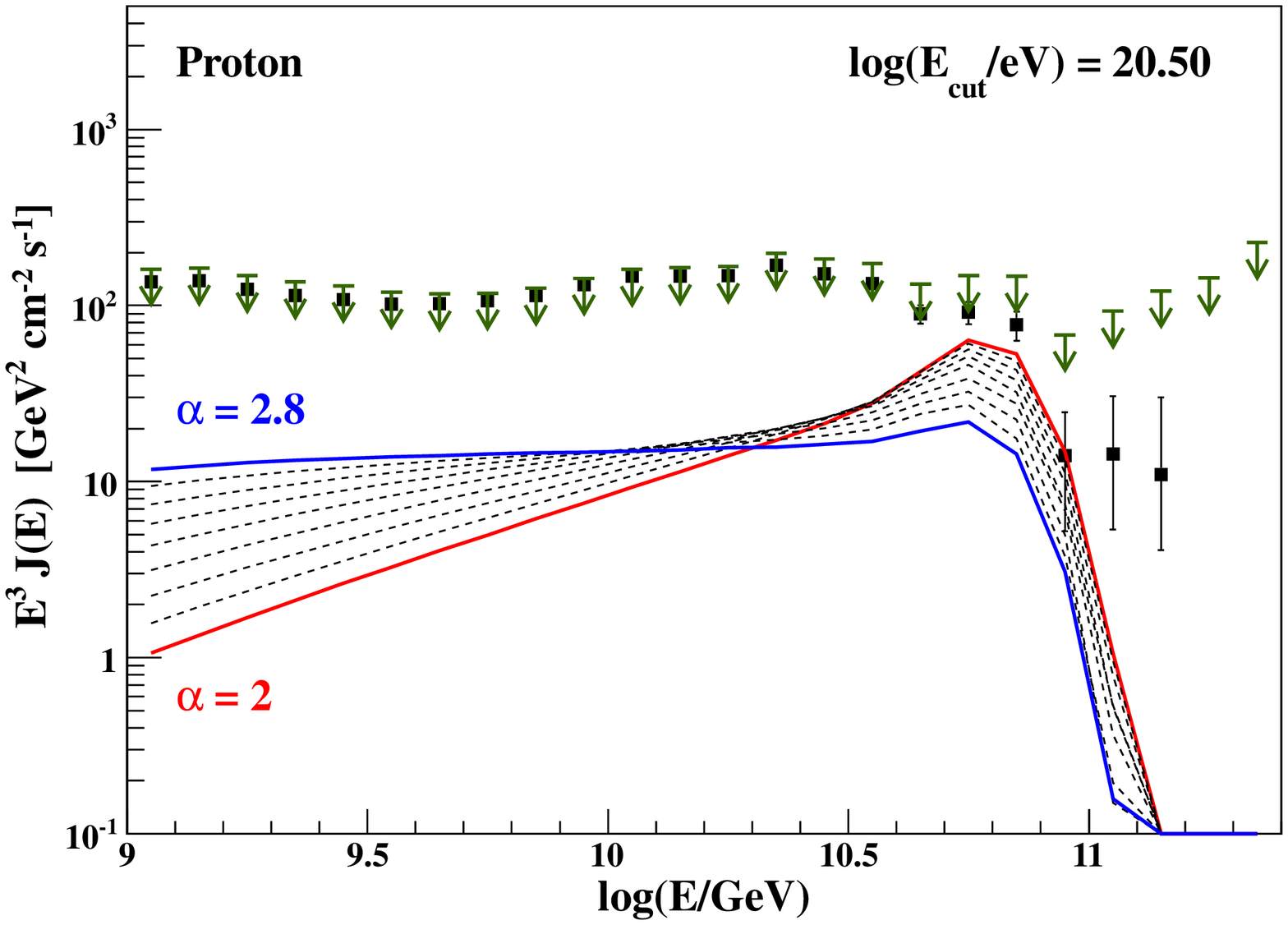}}
  \subfloat[Fe $- E_{cut} = 26\times10^{20.5}$ eV]{\includegraphics[width=0.5\textwidth]{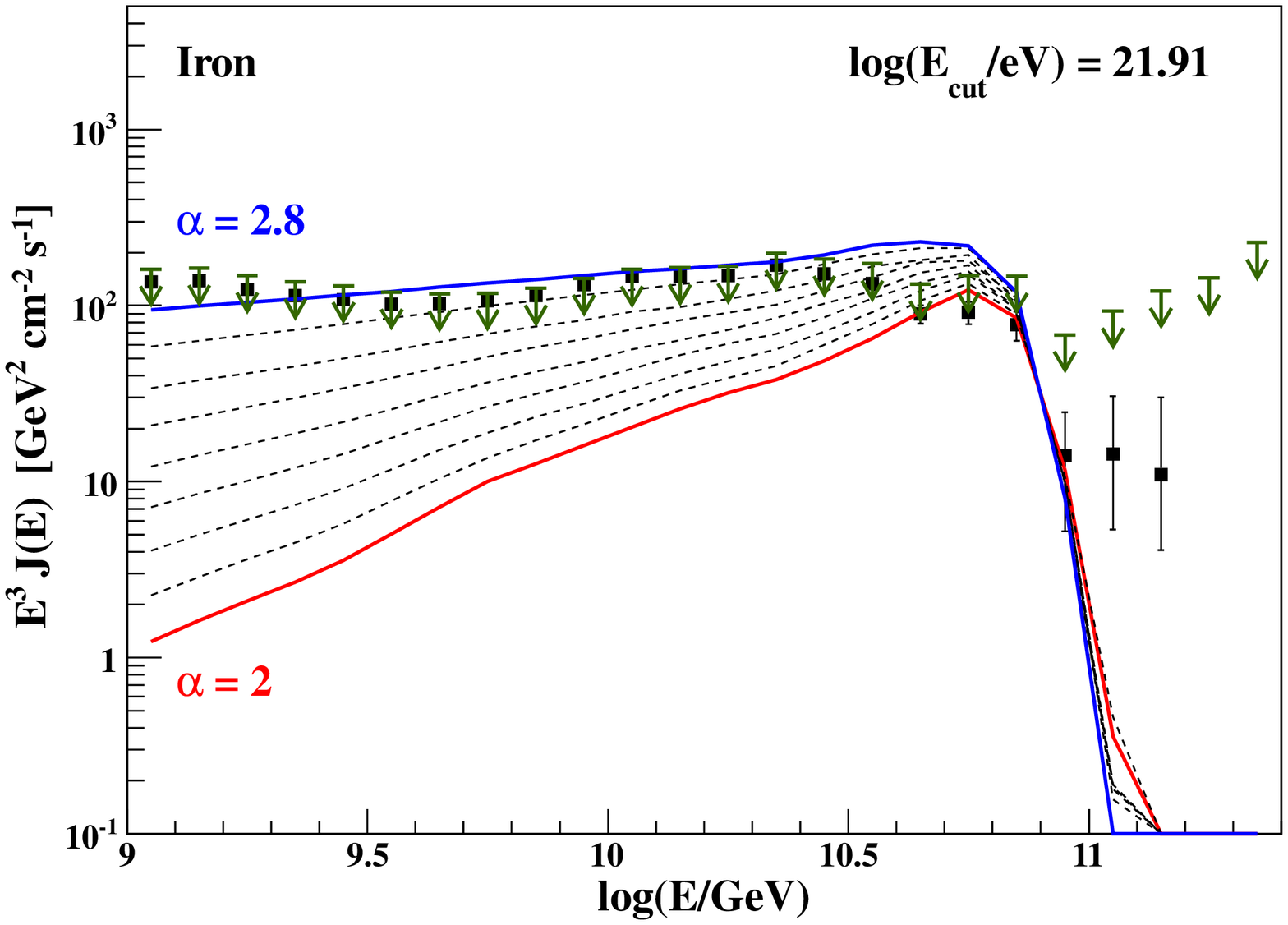}}\\

  \subfloat[Pr $- E_{cut} = 10^{21}$ eV]{\includegraphics[width=0.5\textwidth]{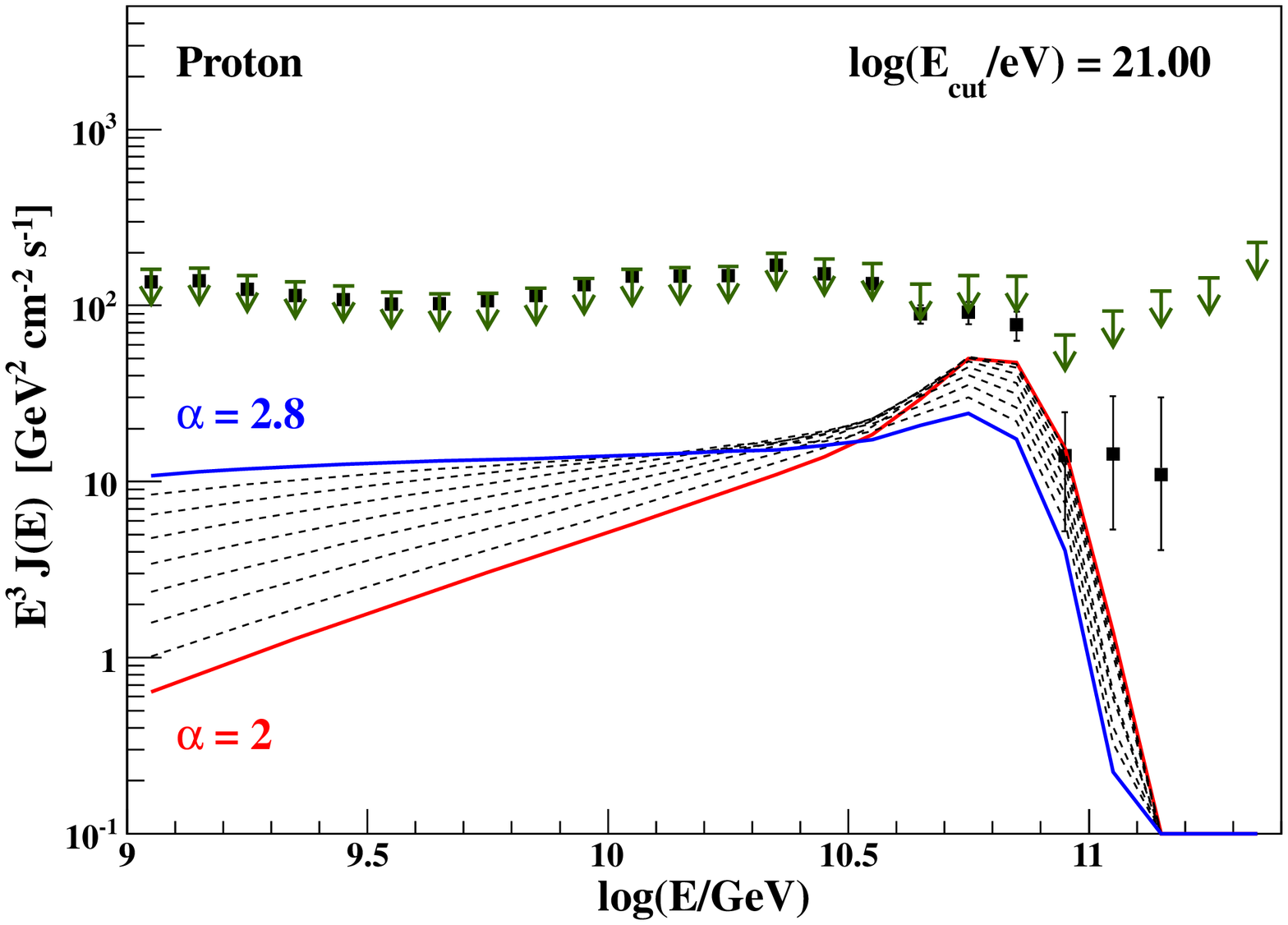}}
  \subfloat[Fe $- E_{cut} = 26\times10^{21}$ eV]{\includegraphics[width=0.5\textwidth]{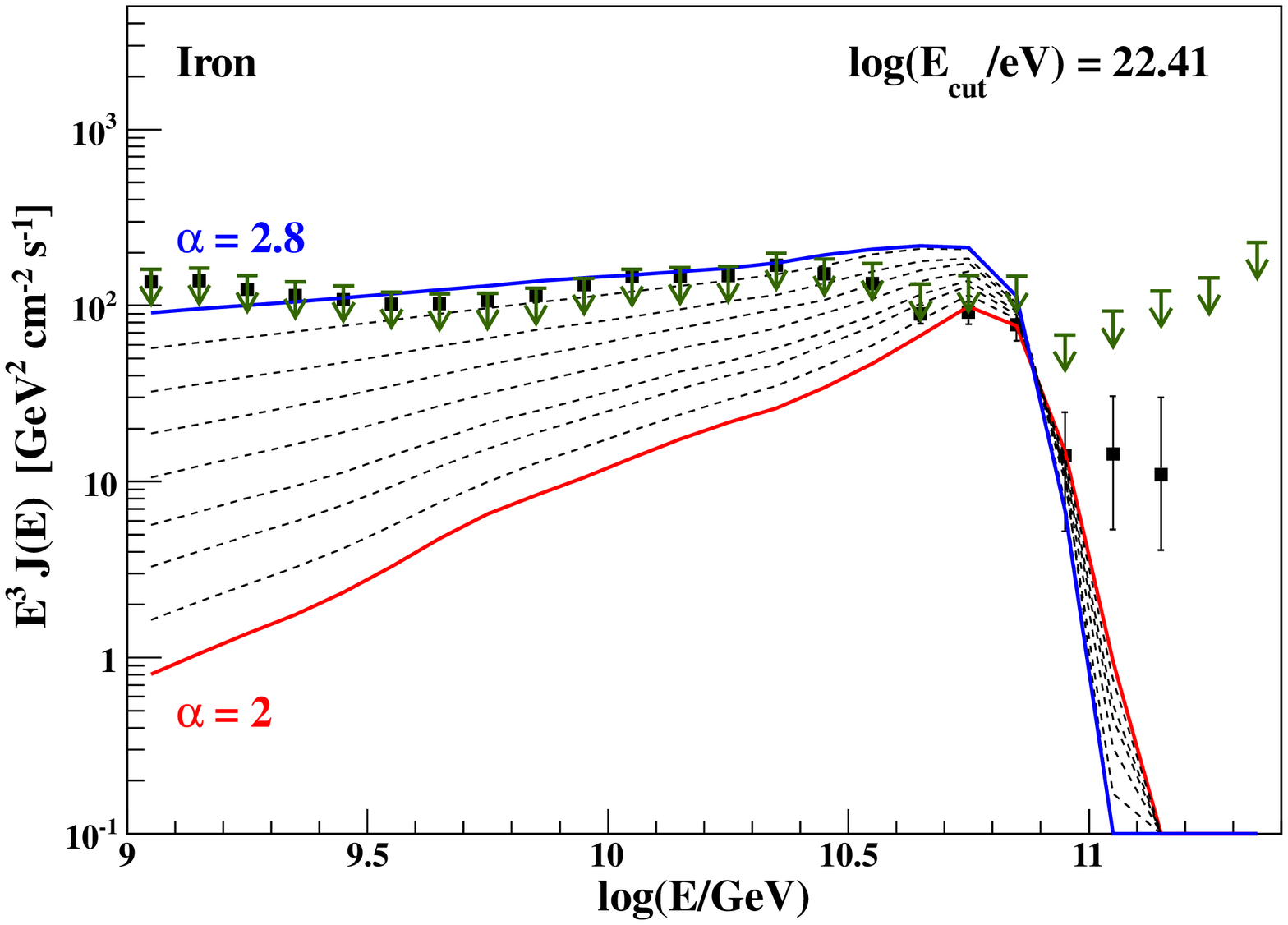}}\\
  \caption{Upper limit on the UHECR cosmic ray flux for Pictor A calculated by using the upper limit on the integral flux of
           GeV-TeV gamma-rays measured by \hess. The points with error bars correspond to the combined spectrum measured by
           the Pierre Auger Observatory and the arrows correspond to the upper limit on the flux, at $99.9\%$ CL, obtained from
           the surface detector data, assuming Poisson statistics. Left column: Figures (a), (c), and (e) correspond to proton
           primaries. Right column: Figures (b), (d), and (f) correspond to iron primaries.}
  \label{fig:pictor:a}
\end{figure}

Figure~\ref{fig:pictor:a} shows the contribution of the upper limit on the UHECR flux of
Pictor A to the total flux observed by the Pierre Auger Observatory. Note that the calculated
flux is weighted by the exposure of the Pierre Auger Observatory in the direction of Pictor A
(see appendix~\ref{app:weight} for details). The plots on the left column show the results
for the cases of only primary protons and on the right column the ones for the cases of
only iron nuclei. Each row in figure~\ref{fig:pictor:a} shows a different cutoff energy of
the cosmic rays in the source, $\log(E_{cut}^{pr}/\textrm{eV}) = 20, \; 20.5$ and $21$ for
protons and $\log(E_{cut}^{26}/\textrm{eV}) \cong 21.41, \; 21.91$ and $22.41$ for iron nuclei
($E_{cut}^{26} = 26 \times E_{pr}$). All plots show the combined cosmic-ray spectrum measured
by the Pierre Auger Observatory~\cite{bib:auger:spectrum} and the upper limits on the total
flux (arrows) at $99.9\%$ CL obtained from the surface detector data only (number of events
in each energy bin and the exposure taken from Ref.~\cite{bib:auger:spectrum}), assuming
Poisson statistics.

From figure~\ref{fig:pictor:a} (right column) it can be seen that
for several combinations of cutoff energy and spectral index
and in many energy bins, the contribution of the upper limit UHECR
flux of Pictor A to the total flux observed by Auger, for the case of
iron nuclei primaries, is above the upper limit on the total flux of
Auger (obtained from the surface detector data). Therefore, the upper
limit on the integral flux of gamma-rays measured by \hess does not
impose a valid upper limit on the cosmic-ray luminosity of Pictor A
for the case in which iron nuclei are considered for the analysis.

\begin{figure}[h]
\centerline{\includegraphics[width=0.8\textwidth]{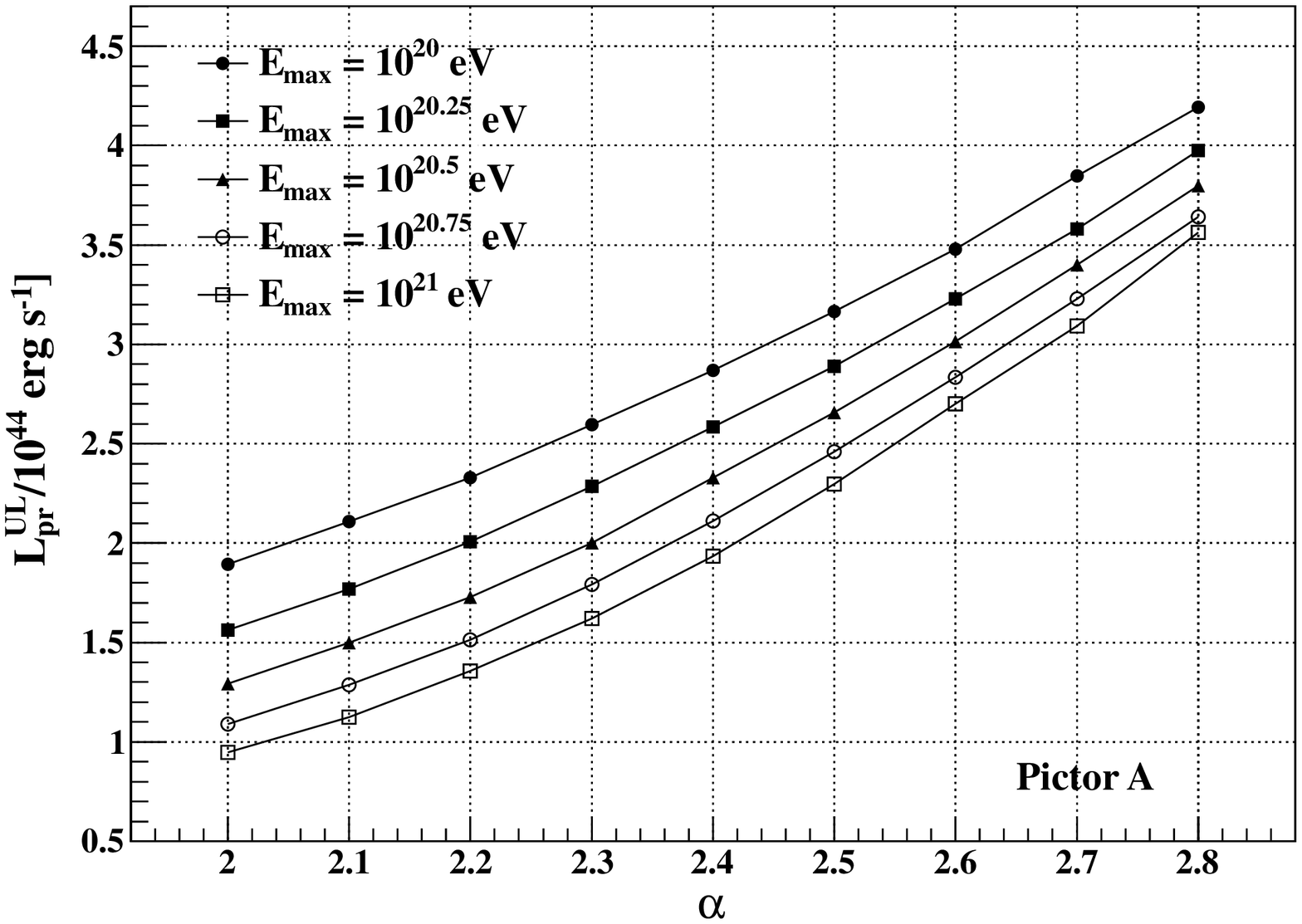}}
\caption{Upper limit on the proton luminosity of Pictor A as a function of the spectral
index of the injection spectrum. The calculation is based on the measured upper limit on
the integral flux of GeV-TeV gamma-rays obtained by \hess. The cutoff energies are:
$\log(E_{cut}/\textrm{eV}) = 20,\ 20.25,\ 20.5,\ 20.75,$ and $21$.}
\label{fig:max:flux:pictor:a}
\end{figure}

On the other hand, the upper limit measured by \hess offers a valid
constrain if UHECR protons are injected by the source. From
figure~\ref{fig:pictor:a} (left column) it can be seen that for all
scenarios considered regarding proton primaries, the contribution of
the upper limit UHECR flux of Pictor A to the total flux observed by
Auger is smaller than the upper limit on the total flux of Auger
(obtained from the surface detector data). In this case, the upper
limit on the integral flux of the gamma-rays determines a valid upper
limit on the proton luminosity of Pictor A.
Figure~\ref{fig:max:flux:pictor:a} shows the upper limit on the proton
luminosity of Pictor A as a function of the spectral index for
$\log(E_{cut}/\textrm{eV}) = 20,\ 20.25,\ 20.5,\ 20.75,$ and $21$.

Pictor A has been observed by Fermi-LAT~\cite{bib:pictor:a:3} in the energy range
from 0.2 to 300 GeV. However the upper limit on the integral gamma-ray flux obtained
by \hess is the most restrictive condition for the proton luminosity of Pictor A. The
flux measured by Fermi-LAT in the energy range from 0.2 to 300 GeV is more than one
order of magnitude larger than the one derived from the upper limit measured by \hess,
independent of spectral index and cutoff energy.

As mentioned before, upper limits on the photon fraction at the highest energies have been obtained
by the past and current cosmic-ray observatories. The most stringent ones come from the observations
done by the Pierre Auger Observatory \cite{bib:AugerPhL:1,bib:AugerPhL:2,bib:AugerPhL:3}. The upper
limits on the integral gamma-ray flux obtained by the Pierre Auger Observatory are more than two
orders of magnitude larger than the integral gamma-ray flux calculated from the upper limit obtained
by \hess, in the same energy range, independent of spectral index and cutoff energy. Therefore, also
in this case the gamma-ray observations at GeV-TeV energies provide the most restrictive upper limit
on the proton luminosity of the source.

\subsection{NGC 7469}
\label{sec:NGC7469}

NGC 7469 is a well known spiral galaxy~\cite{Anderson:70} with a
Seyfert 1 nucleus embedded in a ring of starburst activity. At radio
frequencies, its structure shows an unresolved central component
surrounded by a ring of star forming regions~\cite{Wilson:91}.

\begin{figure}[h]
\centerline{\includegraphics[width=0.8\textwidth]{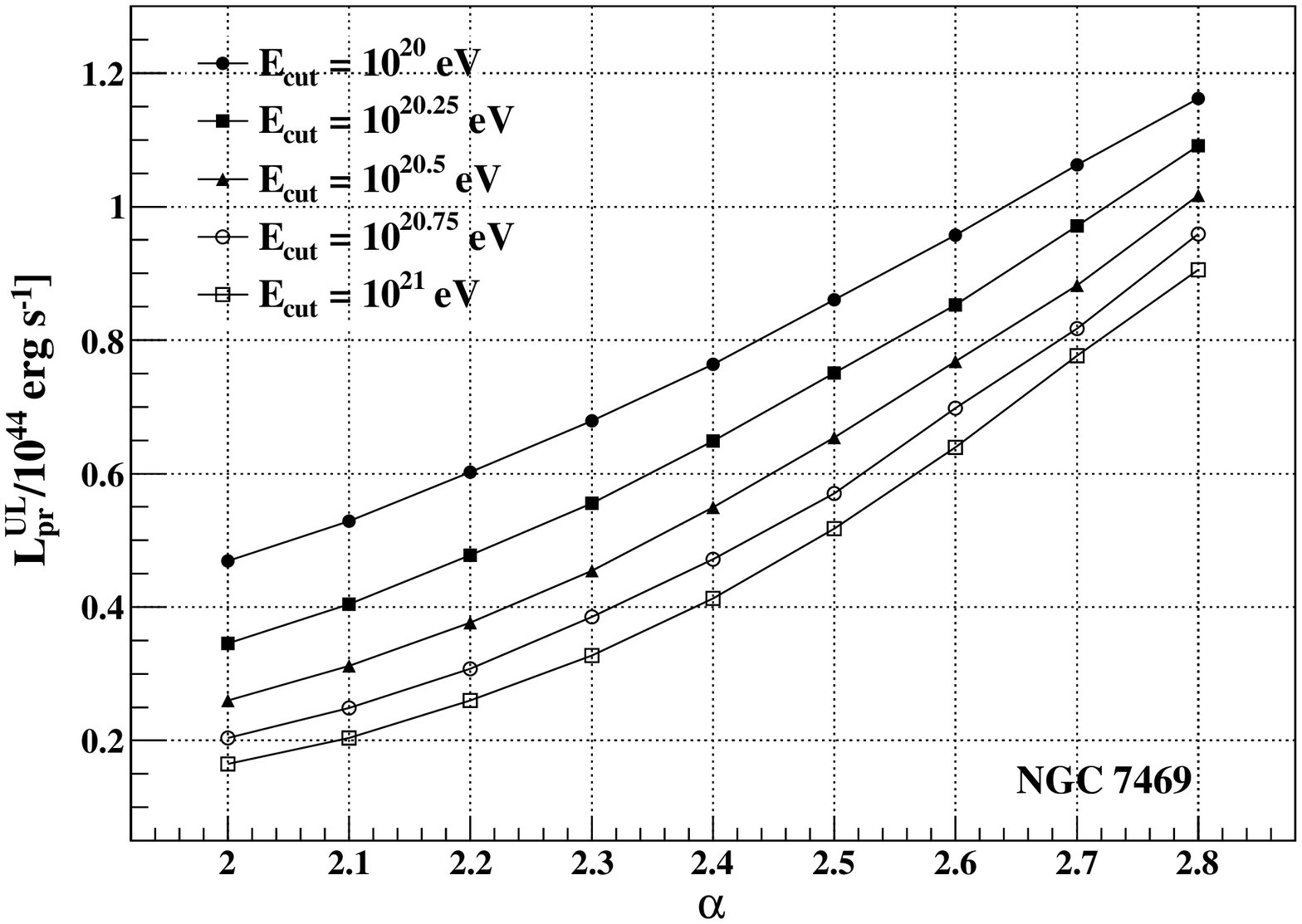}}
\caption{Upper limit on the proton luminosity of NGC 7469 as a function of the spectral
index of the injection spectrum. The calculation is based on the measured upper limit on
the integral flux of GeV-TeV gamma-rays obtained by \hess. The cutoff energies are:
$\log(E_{cut}/\textrm{eV}) = 20,\ 20.25,\ 20.5,\ 20.75,$ and $21$.}
\label{fig:NGC7469}
\end{figure}

The upper limit on the integral gamma-ray flux of NGC 7469 obtained by \hess at 99.9\% CL
is $I_\gamma^{UL}(E>E_\gamma^{th}=330\ \textrm{GeV}) = 1.38 \times 10^{-12}$ cm$^{-2}$ s$^{-1}$.
Also for this source, the upper limit obtained by \hess offers a valid constraint
if UHECR protons are emitted by the source. As for the case of Pictor A, if iron
nuclei are considered the contribution of the upper limit UHECR flux of NGC 7469 to the
total flux measured by Auger is larger than the upper limit UHECR flux of Auger obtained
from the surface detector data. Figures~\ref{fig:NGC7469} shows the upper limits on
the proton luminosity as a function of the spectral index for the same values of the cutoff
energies as the ones considered for Pictor A.

Also for NGC 7469 the upper limits on the integral gamma-ray flux obtained by the Pierre Auger
Observatory, at the highest energies, are more than two orders of magnitude larger than the integral
gamma-ray flux calculated from the upper limit obtained by \hess, in the same energy range, independent
of spectral index and cutoff energy. Therefore, also for this source the gamma-ray observations at
GeV-TeV energies provide the most restrictive upper limit on its proton luminosity.

\section{Conclusions}
\label{sec:conclusion}

In this work a new method to infer an upper limit on the cosmic-ray luminosity of individual
sources from GeV-TeV gamma-ray observations has been presented. This new method is based on
the fact that the UHECR produce gamma-rays when they propagate from the sources on their way
to Earth. The gamma-rays generated in this manner contribute to the total flux observed at
Earth.

Measurements done by ground and space based experiments impose upper limit on the integral
flux of GeV-TeV gamma-rays of some sources. Using a conservative approach of no intrinsic
gamma-ray emission of the source it is possible to calculate an upper limit on the cosmic-ray
luminosity of individual sources.

We have studied in detail the cases of Pictor A and NGC 7469. We have calculated an upper limit
on the proton luminosity of the two sources as a function of the index of the injected spectrum
for several values of the cutoff energies (figures~\ref{fig:max:flux:pictor:a} and \ref{fig:NGC7469}).
We have also shown that the GeV-TeV gamma-ray upper limits obtained by \hess impose the most
restrictive condition on the proton luminosity of the sources.

The results presented in this work illustrate one caveat of the multi-messenger techniques
to study the origin of cosmic rays at the highest energies and gamma-rays at GeV-TeV
energies. The method elaborated here has its motivation increased by the future Cherenkov
Telescope Array which will allow the observation of a much larger number of extragalactic
objects than the ones observed at present with a much lower sensitivity.

\appendix
\section{Simulations}
\label{app:sim}

In this work the propagation of cosmic-ray particles from the source to Earth is simulated
with CRPropa~\cite{bib:crpropa}. CRPropa is one of the most complete public software packages
to model the propagation of nuclei in the intergalactic media considering the most relevant
particle interactions and radiation backgrounds.

CRPropa also propagates the secondary photons generated by the propagation of
the nuclei, which is the main interest of this paper. The most important effect
is the pair production due to the interaction of high energy nuclei with the photon
backgrounds. This effect is implemented as a continuous energy loss according to the
parametrization given in Refs.~\cite{bib:pp:1,bib:pp:2,bib:pp:3}. Other important
effects which produces GeV-TeV gamma rays are also considered in CRPropa, like,
photo-pion production and the subsequent $\pi^0$ decay.

The injection spectrum assumed in this work consists of a power law with an exponential
cutoff (see equation~(\ref{eq:InjSpec})). We have modified CRPropa in order to include
the exponential cutoff.

The propagation of the particles is done with the one dimensional option in CRPropa. The
SOPHIA~\cite{bib:sophia} interaction package, available in CRPropa, is used allowing a
detailed simulation of the photo-pion interaction process. Also, for each studied case
$10^7$ particles are injected at the source position and propagated to the Earth.

Standard cosmology is assumed for the calculations, the Hubble parameter is given by
$H(z) = H_0 \sqrt{\Omega_m (1+z)^3+\Omega_\Lambda}$ with $H_0 = 72$ km s$^{-1}$ Mpc$^{-1}$,
$\Omega_m = 0.26$, and $\Omega_\Lambda = 0.74$.

In the present calculations it is assumed that the Intergalactic
Magnetic Field (IGMF) is negligible. In principle, the presence of a
non-zero IGMF modifies the electromagnetic cascades mainly due to the
deflection of the electrons and positrons in this field. These effects
are negligible if the magnitude of the IGMF is $\lesssim 10^{-14}$ G
\cite{bib:GammaCR:5}. Recently, a lower limit of order of $10^{-15}$ G
has been obtained from gamma-ray observations of distant blazars
\cite{bib:Taylor:11,bib:Neronov:10,bib:Tavecchio:11}. However, in
Ref.~\cite{bib:Arlen:12} it has been suggested that a null IGMF cannot
be excluded due to systematic effects.

\begin{figure}[h]
\centerline{\includegraphics[width=0.8\textwidth]{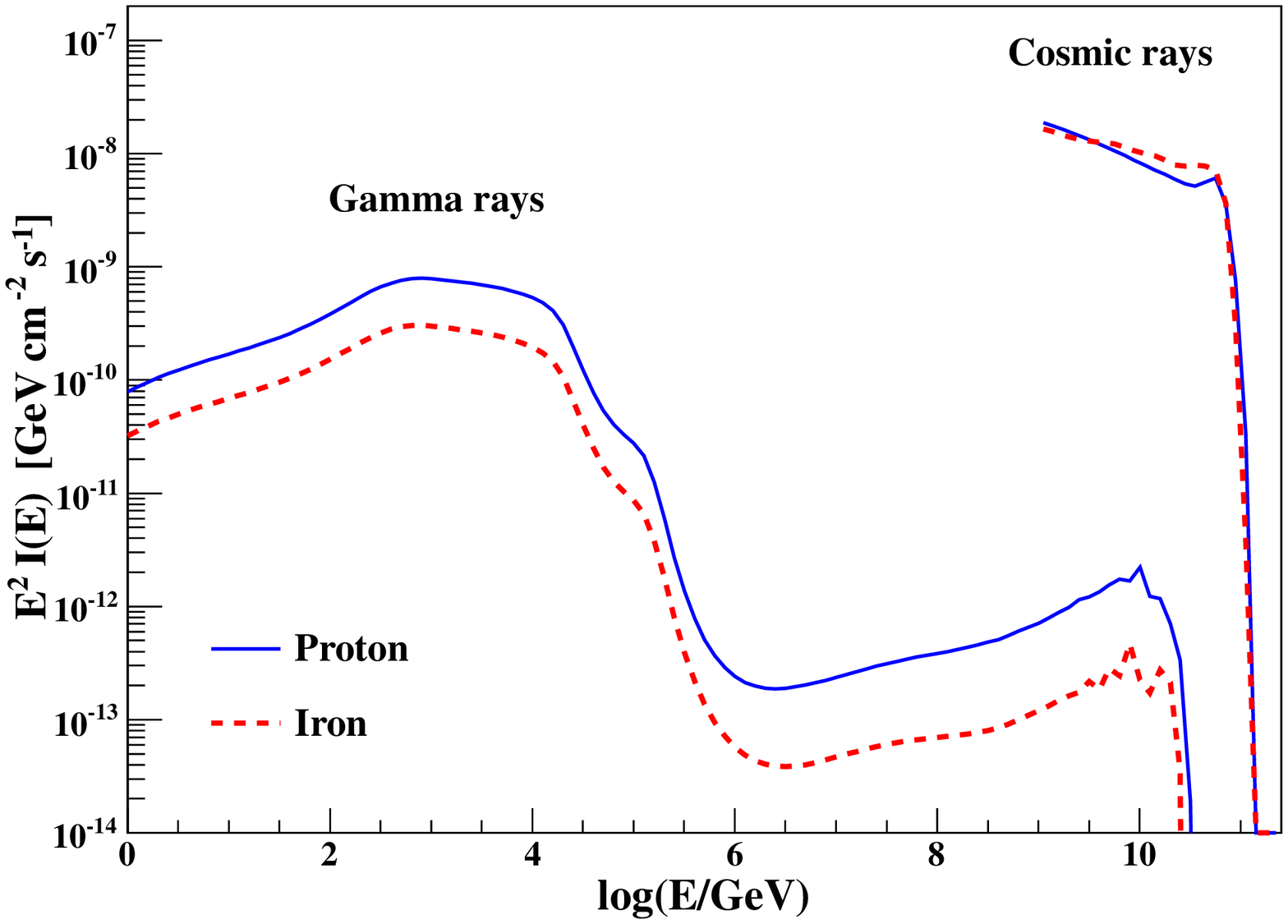}}
\caption{Cosmic-ray and gamma-ray fluxes as a function of the logarithm of the energy
for a source at the distance of Pictor A obtained with CRPropa. Solid lines correspond to
an injection spectra composed by proton and dashed lines to iron nuclei.}
\label{fig:crpropa:ex}
\end{figure}

Figure~\ref{fig:crpropa:ex} illustrates the simulations done with CRPropa. The distance
of Pictor A is used in this example. The spectral index is $\alpha=2.3$ and two different pure
compositions injected by the source are considered, protons and iron nuclei with cutoff energies
$E_{cut}^{p}=10^{20.5}$ eV and $E_{cut}^{26}=26\times10^{20.5}$ eV, respectively. The luminosity
of the source is taken as $L_{CR} = 2.1\times10^{44}$ erg s$^{-1}$. The figure shows the flux at
Earth corresponding to cosmic rays and gamma-rays for the proton and iron nuclei cases. The
propagated spectra shown in this figure are calculated from $K$ and $P$ used in section~\ref{sec:method}.

\section{Contribution of a point source to the measured UHECR flux}
\label{app:weight}

Cosmic ray observatories detect particles coming from all directions
in the sky over large periods of time. Therefore, each individual
cosmic-ray source contributes in a different way to the observed spectrum.
The flux of a given source is weighted by a factor that depends on its
position in the sky, on the location of the observatory, and on the specific
characteristics of the detectors and methods used for the data analysis.

The differential number of detected cosmic-ray particles that come from one
given source is given by
\begin{equation}
\frac{dN_s}{dE} = \Phi_s(E) \; A \; \varepsilon_0 \int_0^{T_{obs}}  dt\ \Theta(\theta_{max}-\theta_s(t))\ \cos \theta_s(t),
\label{eq:df}
\end{equation}
where subscript $s$ is used to specify a given source, $\Phi_s(E)$ is the source flux at Earth,
$A$ is the area of the observatory, $T_{obs}$ is the observation time, $\theta_s$ is the source
zenith angle, and $\varepsilon_0$ is the efficiency of the observatory. Usually $\varepsilon_0$
depends on primary energy and zenith angle, however $\varepsilon_0$ is taken as a constant which
is true for the Pierre Auger Observatory for primary energies $E \gtrsim 10^{18.5}$ eV and zenith
angles $\theta \lesssim 60^\circ$~\cite{bib:auger:trigger} and also for the Telescope Array for
$E \gtrsim 10^{18.8}$ eV and $\theta \lesssim 45^\circ$~\cite{Ivanov:12}. Here $\Theta(x)$ is the
Heaviside function ($\Theta(x)=1$ for $x>0$ and $\Theta(x)=0$ otherwise) and it is assumed that
the events used to calculate the measured spectrum are such that $\theta \le \theta_{max}$.

The observed cosmic-ray flux ($J_{CR}^s(E)$) is obtained dividing the differential number of events
by the exposure ($\mathcal{E}$)

\begin{equation}
J_{CR}^s(E) =\frac{1}{\mathcal{E}} \frac{dN_s}{dE},
\label{eq:jcr}
\end{equation}
and for a constant efficiency the exposure can be written as
\begin{equation}
\mathcal{E}(E) = T_{obs}\ A\ \varepsilon_0\ \pi\ \sin^2\theta_{max}\ \Theta(E-E_0),
\label{eq:exp}
\end{equation}
where $E_0$ is the threshold energy used to calculate the energy spectrum.

Combining equations~(\ref{eq:df}), (\ref{eq:jcr}), and (\ref{eq:exp}) the contribution of a point
source to the total flux can be written as
\begin{equation}
J_{CR}^s(E) =  W_s \Phi_s(E),
\label{eq:jcr2}
\end{equation}
where $W_s =  {\omega_s} / { (\pi\ \sin^2\theta_{max}) } $ and
\begin{equation}
\omega_s = \frac{1}{T_{obs}} \int_0^{T_{obs}}  dt\ \Theta(\theta_{max}-\theta_s(t))\ \cos \theta_s(t).
\label{eq:w}
\end{equation}
For an analytical expression of $\omega_s$ see Ref.~\cite{bib:Sommers:01}.

\acknowledgments
VdS thank the support of the Brazilian population via CNPq and FAPESP
(2010/19514-6). VdS is also in debt with Edivaldo Moura Santos and Rita de C\'assia dos Anjos.
ADS is member of the Carrera del Investigador Cient\'ifico of CONICET, Argentina.
The work of ADS is supported by CONICET PIP 114-201101-00360 and ANPCyT PICT-2011-2223,
Argentina.
The authors thank the Pierre Auger Collaboration for permission to use their
data prior to journal publication.


\begin{thebibliography}{99}

\bibitem{bib:Auger:CenA} The Pierre Auger Collaboration, \emph{Update on the correlation of
the highest energy cosmic rays with nearby extragalactic matter},
\emph{Astropart. Phys.}, {\bf 34} (2010) 314-326.

\bibitem{bib:UpperLim} Fukushima, M., \emph{Measurement of Ultra-High Energy Cosmic Rays: An
Experimental Summary and Prospects}, \emph{Proceedings of the symposium UHECR-2012 at CERN},
arXiv:1302.5893.

\bibitem{bib:Fermi} The Fermi-LAT Collaboration, \emph{Fermi Large Area Telescope Second Source Catalog},
\emph{Astrophys. J. Suppl.}, {\bf 199} (2012) 31.

\bibitem{bib:GammaCR:1} Gabici, S. and Aharonian, F., \emph{Pointlike Gamma Ray Sources as Signatures
of Distant Accelerators of Ultrahigh Energy Cosmic Rays}, \emph{Phys. Rev. Lett.}, {\bf 95} (2005) 251102.

\bibitem{bib:GammaCR:2} Armengaud, E., Sigl, G., and Miniati, F., \emph{Gamma Ray Astronomy with Magnetized
Zevatrons}, arXiv:astro-ph/0511277.

\bibitem{bib:GammaCR:3} Ferrignoa, C., Blasi, P., and De Marco, D., \emph{High energy gamma ray counterparts
of astrophysical sources of ultra-high energy cosmic rays}, \emph{Astropart. Phys.}, {\bf 23} (2005) 211-226.

\bibitem{bib:GammaCR:4} Kotera, K., Allard, D., and Lemoine, M., \emph{Detectability of ultrahigh
energy cosmic-ray signatures in gamma-rays}, {\emph Astronomy and Astrophysics}, {\bf 527} (2011) A54.

\bibitem{bib:GammaCR:5} Ahlers, M. and Salvado, J., \emph{Cosmogenic gamma rays and the composition of
cosmic rays}, \emph{Phys. Rev. D}, {\bf 84} (2011) 085019.

\bibitem{bib:pp:1}
Puget, J., Stecker, F., and Bredekamp, J., \emph{Photonuclear Interactions of Ultrahigh-Energy Cosmic Rays
and their Astrophysical Consequences}, \emph{Astrophys. J.}, {\bf 205} (1976) 638.

\bibitem{bib:pp:2}
Blumenthal, G., \emph{Energy loss of high-energy cosmic rays in
pair-producing collisions with ambient photons}, \emph{Phys. Rev. D}, {\bf 1} (1970) 1596.

\bibitem{bib:pp:3}
  Kelner, S., Aharonian, F., \emph{Energy spectra of gamma rays,
    electrons, and neutrinos produced at interactions of relativistic
    protons with low energy radiation}, \emph{Physical Review D}, {\bf
    78} (2008) 034013.

\bibitem{bib:DeAngelies:13} De Angelis, A., Galanti, G., and Roncadelli, M., \emph{Transparency of
the Universe to gamma rays}, \emph{Monthly Notices of the Royal Astronomical Society}, {\bf 432}
(2013) 3245.


\bibitem{bib:GammaPr:1} Essey, W., Kalashev, O., Kusenko, A., and Beacom, J., \emph{Secondary
Photons and Neutrinos from Cosmic Rays Produced by Distant Blazars}, \emph{Phys. Rev. Lett.},
{\bf 104} (2010) 141102.

\bibitem{bib:GammaPr:2} Essey, W. and Kusenko, A., \emph{A new interpretation of the gamma-ray
observations of distant active galactic nuclei}, \emph{Astropart. Phys.}, {\bf 33} (2010)
81-85.

\bibitem{bib:GammaPr:3} Essey, W., Kalashev, O., Kusenko, A., and Beacom, J., \emph{Role of Line-of-sight
Cosmic-ray Interactions in Forming the Spectra of Distant Blazars in TeV Gamma Rays and High-energy Neutrinos},
{\emph Astrophys. J.}, {\bf 731} (2011) 51.

\bibitem{bib:hess}
\hess Collaboration, \emph{Observations of the Crab nebula with HESS},
\emph{Astronomy and Astrophysics}, {\bf 457} (2006) 899.

\bibitem{bib:pictor:a:chandra}
Wilson, A. S., Young, A. J., and Shopbell, P. L., \emph{Chandra
  X-Ray Observations of Pictor A: High-Energy Cosmic Rays in a Radio
  Galaxy},  \emph{ApJ}, {\bf 547} (2001) 740.

\bibitem{bib:cta}
  CTA Consortium,\emph{Design Concepts for the Cherenkov Telescope
    Array}, \emph{Experimental Astronomy}, {\bf 32} (2011) 193-316.

\bibitem{bib:jem:euso} Takahashi, Y. and the JEM-EUSO Collaboration, \emph{The JEM-EUSO mision},
\emph{New Journal of Physics}, {\bf 11} (2009) 065009.

\bibitem{bib:auger:next} The Pierre Auger Collaboration, \emph{The Next Frontier
in UHECR Research with an Upgraded Pierre Auger Observatory}, arXiv:1307.0226.


\bibitem{bib:multi:1}
Becker, J., \emph{High-energy neutrinos in the context of
  multimessenger physics}, \emph{Physics Report} {\bf 458} (2008) 173.

\bibitem{bib:multi:2}
Allard, D., \emph{Extragalactic propagation of ultrahigh energy
cosmic-rays}, \emph{Astropart. Phys.}, {\bf 39-40} (2012) 33.

\bibitem{bib:multi:3}
Kachelriess, M., Ostapchenko, S., Tomas, R., \emph{Multi-messenger
astronomy with Centaurus A}, \emph{Int. J. Mod. Phys. D}, {\bf 18} (2009)
1591.

\bibitem{bib:crpropa}
Kampert, K., et al., \emph{CRPropa 2.0 - a Public Framework for
Propagating High Energy Nuclei, Secondary Gamma Rays and Neutrinos},
\emph{Astopart. Phys.}, {\bf 42} (2013) 41.

\bibitem{bib:hess:meas}
  Aharonian, F. et al, \emph{Upper limits from HESS active galactic
    nuclei observations in 2005-2007},  \emph {Astronomy and Astrophysics},
  {\bf 478} (2008) 387.

\bibitem{Ivanov:12} Ivanov, D., \emph{Energy spectrum measured by the Telescope Array Surface Detector},  \emph{PhD thesis 2012},
available at http://www.telescopearray.org/index.php/research/publications/theses.

\bibitem{bib:pictor:a:1}
Eracleous, M. and Halpern, J., \emph{Accurate Redshifts and
  Classifications for 110 Radio-Loud Active Galactic Nuclei},
\emph{The Astrophysical Journal Supplement Series}, {\bf 150} (2004) 181.

\bibitem{bib:pictor:a:2}
Liu, F. and Zhang, Y., \emph{A new list of extra-galactic radio jets},\emph {Astronomy\&Astrophysics}, {\bf 381}
(2002) 757.

\bibitem{bib:pictor:a:3}
Anthony, B. and  Adams, J., \emph{Discovery of gamma-ray emission from the Broad Line Radio
Galaxy Pictor A}, \emph{Monthly Notices of the Royal Astronomical Society}, {\bf 421} (2012) 2303.


\bibitem{bib:pictor:a:review:1}
Tingay, S., Lenc, E., Brunetti, G., and Bondi, M., \emph{A high
  resolution view of the jet termination shock in a hot spot of the
  nearby radio galaxy Pictor A: implications for X-ray models of radio
  galaxy hot spots},\emph{The Astronomical Journal}, {\bf 136} (2008)
2473.

\bibitem{bib:pictor:a:4}
Perley, R., R\"oser, H., and Meisenheimer, K., \emph{The radio galaxy Pictor A - a study with the VLA},
\emph{Astron. Astrophys.}, {\bf 328} (1997) 12.

\bibitem{bib:hillas:criteria}  Hillas, A. M. \emph{Cosmic Rays: Recent
Progress and some Current Questions}, (2006) astro-ph:0607109.

\bibitem{bib:lovelace} Lovelace, R. V. E., \emph{Dynamo model of double radio sources}, \emph{Nature}, {\bf
262} (1976) 649.

\bibitem{bib:auger:spectrum}
    Salamida, F. for The Pierre Auger Collaboration, \emph{Update on the
    measurement of the CR energy spectrum above $10^{18}$ eV made using the
    Pierre Auger Observatory}, \emph{Proceedings of the 32nd International Cosmic Ray
    Conference, Beijing, China, 2011}.

\bibitem{bib:AugerPhL:2} The Pierre Auger Collaboration, \emph{Upper limit on the cosmic-ray
photon flux above $10^{19}$ eV using the surface detector of the Pierre Auger Observatory},
\emph{Astropart. Phys.}, {\bf 29} (2008) 243.

\bibitem{bib:AugerPhL:3} Settimo, M. for The Pierre Auger Collaboration, \emph{An update on a search
for ultra-high energy photons using the Pierre Auger Observatory}, \emph{Proceedings of 32nd International
Cosmic Ray Conference, Beijing, China, 2011}, arXiv:1107.4805.

\bibitem{bib:AugerPhL:1} The Pierre Auger Collaboration, \emph{An upper limit to the photon
fraction in cosmic rays above $10^{19}$ eV from the Pierre Auger Observatory}, \emph{Astropart. Phys.},
{\bf 27} (2007) 155.

\bibitem{Anderson:70} Anderson, K., \emph{Spectrophotometry of Eight Seyfert Galaxies}, \emph{Astrophysical Journal},
{\bf 162} (1970) 743.

\bibitem{Wilson:91} Wilson, A., Helfer, T., Haniff, H., and Ward, M., \emph{The starburst ring around the Seyfert nucleus in NGC 7469},
\emph{Astrophysical Journal}, {\bf 381} (1991) 79.

\bibitem{bib:sophia}
M\"ucke, A., Engel, R., Rachen, J., Protheroe, R., and Stanev, T.,
\emph{Monte Carlo simulations of photohadronic processes in astrophysics}, \emph{Comp. Phys. Commun.}, {\bf 124} (2001) 290.

\bibitem{bib:Taylor:11} Taylor, A., Vovk, I., and Neronov, A., \emph{EGMF Constraints from Simultaneous
GeV-TeV Observations of Blazars}, \emph{Astronomy and Astrophysics}, {\bf 529} (2011) A114.

\bibitem{bib:Neronov:10} Neronov, A. and Vovk, I., \emph{Evidence for strong extragalactic magnetic
fields from Fermi observations of TeV blazars}, \emph{Science}, {\bf 328} (2010) 73.

\bibitem{bib:Tavecchio:11} Tavecchio, F., et al., \emph{Extreme TeV blazars and the intergalactic magnetic field},
\emph{Monthly Notices of the Royal Astronomical Society}, {\bf 414} (2011) 3566.

\bibitem{bib:Arlen:12} Arlen, T., et al., \emph{Intergalactic Magnetic Fields and Gamma Ray Observations of
Extreme TeV Blazars}, arXiv:1210.2802.

\bibitem{bib:auger:trigger}
The Pierre Auger Collaboration, \emph{Trigger and aperture of the surface detector array of the
Pierre Auger Observatory}, \emph{Nuclear Instruments and Methods in
Physics Research A}, {\bf 13} (2010) 29.

\bibitem{bib:Sommers:01} Sommers, P., \emph{Cosmic ray anisotropy analysis with a full-sky observatory},
\emph{Astropart. Phys.}, {\bf 14} (2001) 271.

\end{thebibliography}
\end{document}